# Metamaterials and Negative Refractive Index

D. R. Smith[1*], J. B. Pendry[2*], and M. C. K. Wiltshire[2*]

**Recently, artificially constructed metamaterials have become of considerable interest, as these materials can exhibit electromagnetic characteristics unlike any conventional materials. Artificial magnetism and negative refractive index are two specific types of behavior that have been demonstrated over the past few years, illustrating the new physics and new applications possible when we expand our view as to what constitutes a material. In this review, we describe recent advances in metamaterials research, and discuss the potential that these materials may hold for realizing new and seemingly exotic electromagnetic phenomena.**

Consider light passing through a plate of glass. We know that light is an electromagnetic wave, consisting of oscillating electric and magnetic fields, and characterized by a wavelength, $\lambda$. Because visible light has a wavelength that is hundreds of times larger than the atoms of which the glass is composed, the atomic details lose importance in describing how the glass interacts with light. In practice, we can average over the atomic scale, conceptually replacing the otherwise inhomogeneous medium by a homogeneous material characterized by just two macroscopic electromagnetic parameters: The electric permittivity $\varepsilon$ and the magnetic permeability $\mu$.

From the electromagnetic point of view, the wavelength, $\lambda$, determines whether a collection of atoms or other objects can be considered a material. The electromagnetic parameters $\varepsilon$ and $\mu$ need not arise strictly from the response of atoms or molecules: any collection of objects whose size and spacing are much smaller than $\lambda$ can be described by an $\varepsilon$ and $\mu$. Here, the values of $\varepsilon$ and $\mu$ are determined by the scattering properties of the structured objects. While such an inhomogeneous collection may not satisfy our intuitive definition of a material, an electromagnetic wave passing through the structure cannot tell the difference. From the electromagnetic point-of-view, we have created an artificial material—or metamaterial.

The engineered response of metamaterials has had a dramatic impact on the physics, optics and engineering communities, as metamaterials can offer electromagnetic properties that are difficult or impossible to achieve with conventional, naturally occurring materials. The advent of metamaterials has yielded new opportunities to realize physical phenomena that were previously only theoretical exercises.

## Artificial Magnetism

In 1999, several artificial materials were introduced, based on conducting elements designed to provide a magnetic response at microwave and lower frequencies (1). These nonmagnetic structures consisted of arrays of wire loops in which an external applied magnetic field could induce a current, thus producing a magnetic response.

The possibility of magnetism without inherently magnetic materials turns out to be a natural match for Magnetic Resonance Imaging (MRI), which serves as our first example. In an MRI machine there are two distinct magnetic fields. Large quasi-static fields, between 0.2 and 3 tesla in commercial machines, cause the nuclear spins in a patient's body to align. The spins are resonant at the local Larmor frequency typically between 8.5 and 128 MHz, so that a second magnetic field in the form of a radio frequency (RF) pulse will excite them, causing them to precess about the main field. Images are reconstructed by observing the time dependent signal resulting from the precession of the spins. Although the resolution of an MRI machine is obtained through the quasi-static fields, precise control of the RF field is also vital to the efficient and accurate operation of the machine.

Any material destined for use in the MRI environment must not perturb the quasi static magnetic field pattern, thus excluding the use of all conventional magnetic materials. However magnetic metamaterials that respond to time varying fields but not to static fields can be used to alter and focus the RF fields without interfering with the quasi-static field pattern.

All measurements are made on a length scale much smaller than a wavelength, which is 15 m at 20MHz. On a sub-wavelength scale the electric and magnetic components of electromagnetic radiation are essentially independent; so to manipulate a magnetic signal at RF, we need only control the permeability of the metamaterial: the dielectric properties are largely irrelevant.

The metamaterial design best suited to MRI applications is the so-called "Swiss Roll" (1) manufactured by rolling an insulated metallic sheet around a cylinder. A design with about 30 turns on a 1cm diameter cylinder gives a resonant response at 21MHz. Figure 1A shows one such cylinder. The metamaterial is formed by stacking together many of these cylinders.

In an early demonstration, it was shown that Swiss Roll metamaterials could be applied in the MRI environment (2). A bundle of Swiss rolls was used to duct flux

Department of Physics, University of California, San Diego, 9500 Gilman Drive, La Jolla, CA 92093-0319, USA. [2]Department of Physics, Imperial College London, London SW7 2AZ, UK. [3]Imaging Sciences Department, Imperial College London, Hammersmith Hospital, Du Cane Road, London W12 0HS, UK.
[*]to whom correspondence should be addressed.
Email:
drs@physics.ucsd.edu,
j.pendry@imperial.ac.uk,
michael.wiltshire@imperial.ac.uk

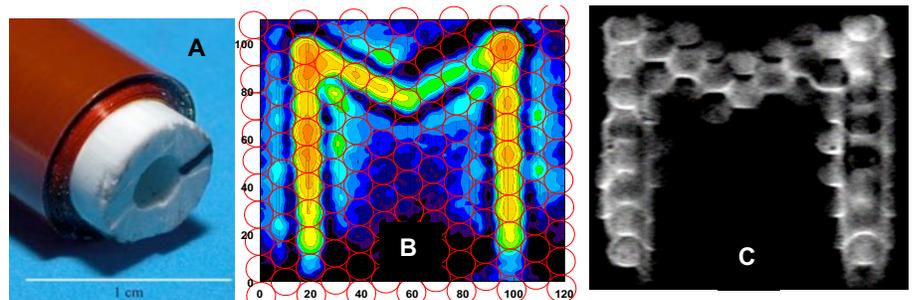

**Fig. 1** (A) A single element of Swiss Roll metamaterial. An array of such elements is assembled into a slab and the RF magnetic field from an M-shaped antenna, placed below the slab, is reproduced on the upper surface. (B) The red circles show the location of the rolls, which were 1 cm in diameter. (C) The resulting image taken in an MRI machine, showing that the field pattern is transmitted back and forth through the slab.

from an object to a remote detector. The metamaterial used in these experiments was lossy, and all the positional information in the image was provided by the spatial encoding system of the MRI machine. Nevertheless, it was clear from this work that such metamaterials could perform a potentially useful and unique function.

## Metamaterials and Resonant Response

Why does a set of conductors shaped into Swiss Rolls behave like a magnetic material? In this structure the coiled copper sheets have a self-capacitance and self-inductance that creates a resonance. The currents which flow when this resonance is activated couple strongly to an applied magnetic field, yielding an effective permeability that can reach quite high values. At the resonant frequency a slab of Swiss Roll composite behaves as a collection of "magnetic wires": A magnetic field distribution incident on one face of the slab is transported uniformly to the other, in the same way that an electric field distribution would be transported by a bundle of electrically conducting wires. In real materials, of course, there is loss, and this limits the resolution of the transfer to roughly $d/\sqrt{\text{Im}(\mu)}$ where $d$ is the thickness of the slab, and $\text{Im}(\mu)$ is the imaginary part of the permeability. This field transference was demonstrated (3) by arranging an antenna in the shape of the letter M as the source, and mapping the transmitted magnetic field distribution (the fields near a current carrying wire are predominantly magnetic). On resonance, the Swiss Roll structure transmitted the incident field pattern across the slab (Fig. 1B), and the resolution matched that predicted by theory.

The image transference was also demonstrated in an MRI machine (4). Here, the same M-shaped antenna was used both as the source of the RF excitation field and as the detector for the signal (see Fig. 1C), and the metamaterial was tested twice over. First, it had to transmit the excitation field without degradation of spatial information, so that the required spin pattern was excited in the sample. Second, the signal from that spin pattern had to be conveyed faithfully back to the receiver. This experiment demonstrated that a high-performance metamaterial could act as a magnetic face-plate, and convey information from one side to the other without loss of spatial information (see Fig. 1C).

Medical imaging is but one example of the potential utility of artificial magnetic materials. Although this artificial magnetism is unique, at these lower frequencies magnetism is also exhibited by existing conventional materials. As we look to higher frequencies, on the other hand, conventional magnetism tails off and artificial magnetism may play an increasingly important role.

A frequency range of particular interest occurs between 1 - 3 THz, a region that represents a natural breakpoint between magnetic and electric response in conventional materials. At lower frequencies, inherently magnetic materials (those whose magnetism results from unpaired electron spins) can be found that exhibit resonances. At higher frequencies, nearly all materials have electronic resonances that result from lattice vibrations or other mechanisms, and give rise to electric response. The mid THz region represents the point where electric response is dying out from the high frequency end, and magnetic response is dying out from the low frequency end: Here, nature does not provide any strongly dielectric or magnetic materials.

Metamaterials, on the other hand, can be constructed to provide this response. At higher frequencies, the split-ring resonator (SRR)—another conducting structure—can be conveniently utilized to achieve a magnetic response (1). The SRR consists of a planar set of concentric rings, each ring with a gap. Because the SRR is planar, it is easily fabricated by lithographic methods at scales from low frequencies to optical frequencies.

Recently, an SRR composite designed to exhibit a magnetic resonance at THz frequencies has been fabricated (5). The size of the SRRs was on the order of 30 microns—ten times smaller than the 300 micron wavelength at 1 THz. Scattering experiments confirmed that the SRR medium had a magnetic resonance that could be tuned throughout the THz band by slight changes to the geometrical SRR parameters.

Both the Swiss Roll metamaterial and the THz SRR metamaterial illustrate the advantage of developing artificial magnetic response. But metamaterials can take us even further, to materials that have no analog in conventional materials.

## Negative Material Response

A harmonic oscillator has a resonant frequency, at which a small driving force can produce a very large displacement. Think of a mass on a spring: Below the resonant frequency, the mass is displaced in the same direction as the applied force. However, above the resonant frequency, the mass is displaced in a direction opposite to the applied force. Because a material can be modeled as a set of harmonically bound charges, the negative resonance response translates directly to a negative material response, with the applied electric or magnetic field acting on the bound charges corresponding to the force, and the responding dipole moment corresponding to the displacement. A resonance in the material response leads to negative values for $\varepsilon$ or $\mu$ above the resonant frequency.

Nearly all familiar materials, such as glass or water have positive values for both $\varepsilon$ and $\mu$. It is less well recognized that materials are common for which $\varepsilon$ is negative. Many metals—silver and gold, for example, have negative $\varepsilon$ at wavelengths in the visible spectrum. A material having either (but not both) $\varepsilon$ or $\mu$ negative is opaque to electromagnetic radiation.

Light cannot get into a metal, or at least it cannot penetrate very far, but metals are not inert to light. It is possible for light to be trapped at the surface of a metal, and propagate around in a state known as a surface plasmon. These surface states have intriguing properties which are just beginning to be exploited in applications (6).

While material response is fully characterized by the parameters $\varepsilon$ and $\mu$, the optical properties of a transparent material are often more conveniently described by a different parameter, the refractive index, $n$, given by $n = \sqrt{\varepsilon\mu}$. A wave travels more slowly in a medium such as glass or water by a factor $n$. All known transparent materials possess a positive index since $\varepsilon$ and $\mu$ are both positive.

Yet, the allowed range of material response does not preclude us from considering a medium for which both $\varepsilon$ and $\mu$ are negative. More than thirty-five years ago Victor Veselago pondered the properties of just such a medium (7). As this product $\varepsilon\mu$ is positive, taking the square root gives a real number for the index. We thus conclude that materials with negative $\varepsilon$ and $\mu$ are transparent to light.

There is a wealth of well known phenomena associated with electromagnetic wave propagation in materials. All of these phenomena must be re-examined when $\varepsilon$ and $\mu$ are simultaneously negative. For example, the Doppler shift is reversed, with a light source moving towards an observer being down-shifted in frequency. Likewise, the Cherenkov radiation from a charge passing through the material is emitted in the opposite direction to the charge's motion rather than in the forward direction (7).

The origin of this newly predicted behavior can be traced to the distinction between the group velocity, which characterizes the flow of energy, and the phase velocity, which characterizes the movement of the wave fronts. In conventional materials, the group and phase velocities are parallel. By contrast, the group and phase velocities point in opposite directions when $\varepsilon < 0$ and $\mu < 0$ (see Fig. 2).

The reversal of phase and group velocity in a material implies a simply stated but profound consequence: The sign of the refractive index, n, must be taken as negative.

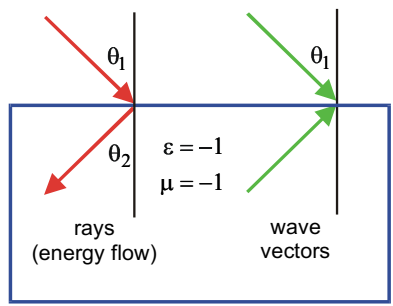

**Fig. 2.** Negative refraction in operation: on the left a ray enters a negatively refracting medium and is bent the 'wrong way' relative to the surface normal forming a chevron at the interface. On the right we sketch the wave vectors: negative refraction requires that the wave vector and group velocity (the ray velocity) point in opposite directions.

After the early work of Veselago, interest in negative index materials evaporated, since no known naturally occurring material exhibits a frequency band with $\mu < 0$ and also possesses $\varepsilon < 0$. The situation changed in 2000, however, when a composite structure based on SRRs was introduced and shown to have a frequency band over which $\varepsilon$ and $\mu$ were both negative (*8*). The negative $\mu$ occurred at frequencies above the resonant frequency of the SRR structure. The negative $\varepsilon$ was induced by interweaving the SRR lattice with a lattice of conducting wires. A lattice of wires possesses a cut-off frequency, below which $\varepsilon$ is negative (*9*); by choosing the parameters of the wire lattice such that the cutoff frequency was significantly above the SRR resonant frequency, the composite was made to have an overlapping region where both $\varepsilon$ and $\mu$ were negative. This preliminary experiment showed that Veselago's hypothesis could be realized in artificial structures, and kicked off the rapidly growing field of negative index metamaterials.

## Negative Refraction and Sub-Wavelength resolution

Experimentally, the refractive index of a material can be determined by measuring the deflection of a beam as it enters or exits the interface to a material at an angle. The quantitative statement of refraction is embodied in Snell's law, which relates the exit angle of a beam, $\theta_2$, as measured with respect to a line drawn perpendicular to the interface of the material, to the angle of incidence, $\theta_1$, by the formula

$$\sin(\theta_2) = n \sin(\theta_1)$$

The refractive index determines the amount by which the beam is deflected. If the index is positive, the exiting beam is deflected to the opposite side of the surface normal, while if the index is negative, the exiting beam is deflected the same side of the normal (see Fig. 2).

In 2001, a Snell's law experiment was performed on a wedge-shaped metamaterial designed to have a negative index of refraction at microwave frequencies (*10*). In this experiment, a beam of microwaves was directed onto the flat portion of the wedge sample, passing through the sample undeflected, and then refracted at the second interface. The angular dependence of the refracted power was then measured around the circumference, establishing the angle of refraction.

The result of the experiment (Fig. 3) indicated quite clearly that the wedge sample refracted the microwave beam in a manner consistent with Snell's law. Fig. 3B shows the detected power as a function of angle for a Teflon wedge ($n = 1.5$, blue curve) versus that of the NIM wedge (red curve). The location of the peak corresponding to the NIM wedge implies an index of −2.7.

While the experimental results appeared to confirm that the metamaterial sample possessed a negative refractive index, the theoretical foundation of negative refraction was challenged in 2002 (*11*). It was argued that the inherent frequency dispersive properties of negative index materials would prevent information carrying signals from truly being negatively refracted. The theoretical issue was subsequently addressed by several authors (*12-14*), who concluded that, indeed, time-varying signals could also be negative refracted.

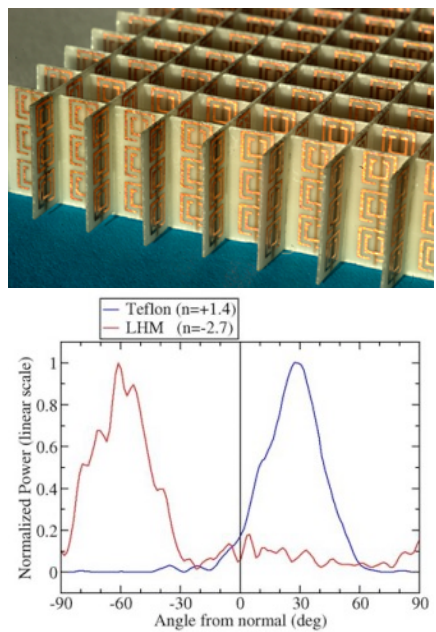

**Fig. 3.** (A) A negative index metamaterial, formed by SRRs and wires deposited on opposite sides lithographically on standard circuit board. The height of the structure is 1 cm. (B) The power detected as a function of angle in a Snell's law experiment performed on a Teflon sample (blue curve) and a negative index sample (red curve).

Since this first demonstration of negative refraction, two more Snell's law experiments have been reported, both using metamaterial wedge samples similar in design to that used in the first demonstration. These experiments have addressed aspects not probed in the first experiment. In one of the experiments, for example, spatial maps of the electromagnetic fields were made as a function of distance from the wedge to the detector. In addition, wedge samples were used with two different surface cuts to confirm that the angle of refraction was consistent with Snell's law (*15*). In the second of these experiments, the negatively refracted beam was measured at much farther distances from the wedge sample (*16*). Moreover, in this latter experiment, the metamaterial sample was carefully designed such that material losses were minimized, and the structure presented a better impedance match to free space; in this manner, much more energy was transmitted through the sample, making the negatively refracted beam easier to observe and much less likely to be the result of any experimental artifacts. These additional measurements have sufficed to convince most that materials with negative refractive index are indeed a reality.

Having established the reality of negative refraction, we are now free to investigate other phenomena related to negative index materials. We quickly find that some of the most long-held notions related to waves and optics must be rethought! A key example is the case of imaging by a lens. It is an accepted convention that the resolution of an image is limited by the wavelength of light employed. The wavelength limitation of optics imposes serious constraints on optical technology: Limits to the density with which DVDs can be written, and the density of electronic circuitry created by lithography are manifestations of the wavelength limitation. Yet, there is no fundamental reason why an image should not be created with arbitrarily high resolution. The wavelength limitation is a result of the optical configuration of conventional imaging.

Negative refraction by a slab of material bends a ray of light back towards the axis and thus has a focusing effect at the point where the refracted rays meet the axis – see figure 4A). It was recently observed (*17*) that a negative index lens exhibits an entirely new type of focusing phenomenon, bringing together not just the propagating rays, but also the finer details of the electromagnetic near fields which are evanescent and do not propagate (see figure 4B). For a planar slab of negative index material, under idealized conditions, an image plane exists that contains a perfect copy of an object placed on the opposite side of the slab. While realizable materials will never meet the idealized conditions, nevertheless these new negative index concepts show that sub-wavelength imaging is possible, in principle; we need no longer dismiss this possibility from consideration.

This trick of including the high resolution but rapidly decaying part of the image is achieved by resonant amplification of the fields. Negative materials support a host of surface modes closely related to surface plasmons, commonly observed at metal surfaces (6), and it is these states that are resonantly excited. By amplifying the decaying fields of a source, the surface modes restore them to the correct amplitude in the image plane.

The term lens is a misnomer when describing focusing by negative materials. Recent work (18, 19) has shown that a more accurate description of a negative index material is negative space. To clarify, imagine a slab of material with thickness $d$ for which,

$$\varepsilon = -1, \quad \mu = -1 \,.$$

Then, optically speaking, it is as if the slab had gained an equal thickness of empty space next to it and annihilated it. In effect the new lens translates an optical object a distance $2d$ down the axis to form an image.

The concept of the perfect lens at first met with considerable opposition (20, 21), but the difficulties raised have been answered by clarifying the concept and its limitations (22, 23), by numerical simulation (24, 25), and in the last few months by experiments.

In a recent experiment, a two dimensional version of a negative index material has been assembled from discrete elements arranged on a planar circuit board (26). A detail of the experiment (Fig. 4C) shows the location of a point source and the expected location of the image. Fig. 4D shows the experimental data, where the red curve is the measured result and lies well within the green curve, the calculated diffraction limited result. A more perfect system with reduced losses would produce better focusing.

The conditions for the "perfect lens" are rather severe and must be met rather accurately (23). This is a particular problem at optical frequencies where any magnetic activity is hard to find. However there is a compromise that we can make if all the dimensions of the system are much less than the wavelength: As stated earlier, over short distances the electric and magnetic fields are independent. We may choose to concentrate entirely on the electric fields in which case it is only necessary to tune to $\varepsilon = -1$ and we can ignore $\mu$ completely. This 'poor mans' lens will focus the electrostatic fields, limited only by losses in the system. Thus it has been proposed that a thin slab of silver a few nanometres thick can act as a lens (17). Experiments have shown amplification of light by such a system in accordance with theoretical predictions (27).

## Photonic Crystals and Negative Refraction

Metamaterials based on conducting elements have been utilized to demonstrate negative refraction with great success. However, the use of conductors at higher frequencies—especially optical—can be problematic due to losses. As an alternative, many researchers have been investigating the potential of negative refraction in the periodic structures known as photonic crystals. These materials are typically composed of insulators, and therefore can exhibit very low losses, even at optical frequencies.

In photonic crystals, the size and periodicity of the scattering elements are on the order of the wavelength, rather than being much smaller. Describing a photonic crystal as a homogeneous medium is inappropriate, so it is not possible to define values of $\varepsilon$ or $\mu$. Nevertheless, diffractive phenomena in photonic crystals can lead to the excitation of waves for which phase and group velocities are reversed, in the same manner as in negative index metamaterials. Thus, under the right conditions, negative refraction can be observed in photonic crystals.

In 2000, it was shown theoretically that several photonic crystal configurations could exhibit the same types of optical phenomena predicted for negative index materials, including negative refraction and imaging by a planar surface (23).

Since then, several versions of photonic crystals have been utilized to demonstrate negative refraction, for example in an experiment on metallic photonic crystal formed into a wedge (29), and in an experiment in which the shift in the exit position of a beam incident at an angle to one face of a flat dielectric photonic crystal slab was measured (30). While these experiments have been performed at microwave frequencies, the same structures scaled to optical frequencies would possess far less loss than the metamaterials based on conducting elements.

Photonic crystals have also been used to demonstrate focusing (31). Images as sharply defined as $\lambda/5$ can be obtained at microwave frequencies using photonic crystal slab lenses (32). The work in photonic crystals is an example of where the identification of new material parameters has prompted the development of similar concepts in other systems.

## New Materials, New Physics

We rely on the electromagnetic material parameters such as the index, or $\varepsilon$ and $\mu$, to replace the complex and irrelevant details of structures much smaller than the wavelength. This is why, for example, we can understand how a lens focuses light without concern about the motion of each of the atoms of which the lens is composed. With the complexity of the material removed from consideration, we are free to use the material properties to design applications, or study other wave propagation phenomena with great flexibility.

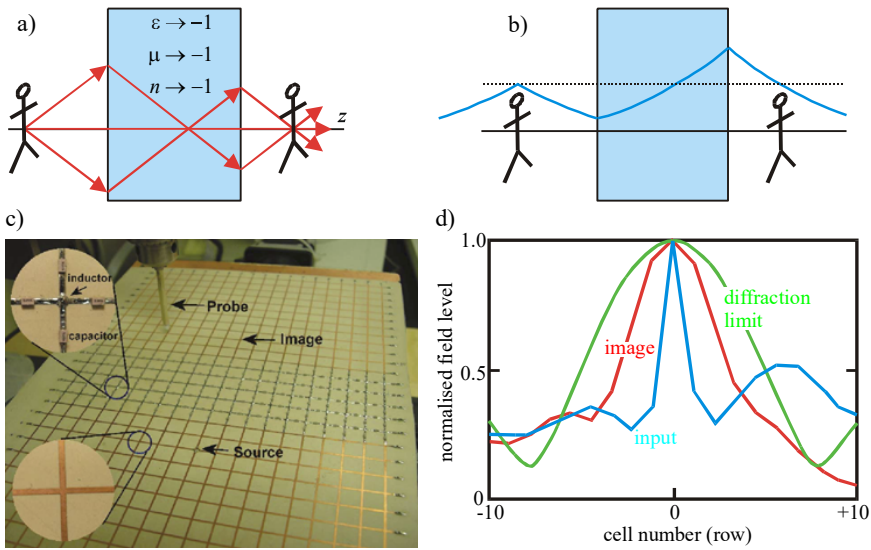

**Fig. 4** Perfect lensing in action: A slab of negative material effectively removes an equal thickness of space for a) the far field and b) the near field translating the object into a perfect image. Microwave experiments by the Eleftheriades group c) demonstrate that sub wavelength focusing is possible, limited only by losses in the system. d) measured data are shown in red and compared to the perfect results shown in blue. Losses limit the resolution to less than perfect, but better than the diffraction limit shown in green.

The past few years have illustrated the power of the metamaterials approach, as new material responses—some with no analog in conventional materials—are now available for exploration. The examples presented here, including artificial magnetism, negative refraction, and near-field focusing are just the earliest of the new phenomena to emerge from the development of artificial materials. As we move forward, our ability to realize the exotic and often dramatic physics predicted for metamaterials will now depend on the quality of metamaterials.

## Acknowledgements

JBP thanks the EC under project FP6-NMP4-CT-2003-505699 for financial support. DRS also acknowledges support from DARPA (DAAD19-00-1-0525), and DRS and JBP for DARPA/ONR MURI grant N00014-01-1-0803. Additionally JBP acknowledges support from the EPSRC, and MCKW from the EC Information Societies Technology (IST) programme Development and Analysis of Left-Handed Materials (DALHM), Project number: IST-2001-35511.

### References

1. J. B. Pendry, A. J. Holden, D. J. Robbins and W. J. Stewart, *IEEE Trans. Micr. Theory and Techniques*, **47**, 2075 (1999).
2. M. C. K. Wiltshire et al., *Science*, **291** (5505), 849 (2001).
3. M. C. K. Wiltshire, J. V. Hajnal, J. B. Pendry, D. J. Edwards, C. J. Stevens, *Optics Express*, **11** (7), 709 (2003).
4. M. C. K. Wiltshire et al. *Proc. Intl. Soc. Mag. Reson. Med.*, **11**, 713 (2003).
5. T. J. Yen et al., *Science*, **303**, 1494 (2004).
6. W. L. Barnes, A. Dereux, T. W. Ebbesen, *Nature*, **424**, 824 (2003).
7. V. G. Veselago, *Soviet Physics USPEKI*, **10**, 509 (1968).
8. D. R. Smith, W. J. Padilla, D. C. Vier, S. C. Nemat-Nasser, S. Schultz, *Phys. Rev. Lett.*, **84**, 4184 (2000).
9. J. B. Pendry, A. J. Holden, W. J. Stewart, I. Youngs, *Phys. Rev. Lett.*, **76**, 4773 (1996).
10. R. Shelby, D. R. Smith, S. Schultz, *Science*, **292**, 77 (2001).
11. P. M. Valanju, R. M. Walser, A. P. Valanju, *Phys. Rev. Lett.*, **88**, 187401 (2002).
12. D. R. Smith, D. Schurig, J. B. Pendry, *Appl. Phys. Lett.*, **81**, 2713 (2002).
13. J. Pacheco, T. M. Grzegorczyk, B.-I. Wu, Y. Zhang, J. A. Kong, *Phys. Rev. Lett.*, **89**, 257401 (2002).
14. J. B. Pendry, D. R. Smith, *Phys. Rev. Lett.*, **90**, 029703 (2003).
15. A. A. Houck, J. B. Brock, I. L Chuang, *Phys. Rev. Lett.*, **90**, 137401 (2003).
16. C. G. Parazzoli, R. B. Greegor, K. Li, B. E. C. Koltenbah, M. Tanielian, *Phys. Rev. Lett.*, **90**, 107401 (2003).
17. J. B. Pendry, *Phys. Rev. Lett.*, **85**, 3966 (2000).
18. J. B. Pendry, S. A. Ramakrishna, *J. Phys.: Condensed Matter*, **15**, 6345 (2003).
19. A. Lakhtakia, *Int. Journal of Infrared and Millimeter Waves*, **23**, 339 (2002).
20. N. Garcia, M. Nieto-Vesperinas, *Phys. Rev. Lett.*, **88**, 207403 (2002).
21. G. W. 't Hooft, *Phys. Rev. Lett.*, **87**, 249701 (2001).
22. J. T. Shen, P. M. Platzman, *Appl. Phys. Lett.*, **80**, 3286 (2002).
23. D. R. Smith, D. Schurig, M. Rosenbluth, S. Schultz, S. A. Ramakrishna, J. B. Pendry, *Appl. Phys. Lett.*, **82**, 1506 (2003).
24. P. Kolinko, D. R. Smith, *Optics Express*, **11**, 640 (2003).
25. S. A. Cummer, *Appl. Phys. Lett.*, **82**, 1503 (2003).
26. A. Grbic, G. V. Eleftheriades, *Phys. Rev. Lett.*, **92**, 117403 (2004).
27. Z. W. Liu, N. Fang, T. J. Yen, X. Zhang, *Appl. Phys. Lett.*, **83**, 5184 (2003).
28. M. Notomi, *Phys. Rev. B*, **62**, 10696 (2000).
29. P. V. Parimi, W. T. Lu, P. Vodo, J. Sokoloff, J. S. Derov, S. Sridhar, *Phys. Rev. Lett.*, **92**, 127401 (2004).
30. E. Cubukcu, K. Aydin, E. Ozbay, S. Foteinopoulou, C. M. Soukoulis, *Nature*, **423**, 604 (2003).
31. P. V. Parimi, W. T. Lu, P. Vodo, S. Sridhar, *Nature*, **426**, 404 (2003).
32. E. Cubukcu, K. Aydin, E. Ozbay, S. Foteinopolou, C. M. Soukoulis, *Phys. Rev. Lett.*, **91**, 207401 (2003).